\documentclass[aps,pra,twocolumn,showpacs,superscriptaddress]{revtex4-1}  
\usepackage{graphicx}
\usepackage{amsmath}
\usepackage{amssymb}
\usepackage{amsfonts}
\usepackage{amsbsy}
\usepackage{color}
\usepackage[colorlinks=true, urlcolor=blue, citecolor=blue, linkcolor=blue]{hyperref}

\renewcommand{\vec}[1]{\mbox{\boldmath$\mathrm{#1}$}}
\let\sb=_ \catcode`\_=\active \def_#1{\ensuremath \sb{\rm#1}}
\def\ind#1{{_{\mathrm{#1}}}}

\usepackage{xcolor}

\begin{document}

\title{{Twisting and tweezing the spin wave:  on vortices, skyrmions,  helical waves, and the magnonic spiral phase plate}}

\author{Chenglong Jia}
\email{cljia@lzu.edu.cn}

\author{Decheng Ma}

\address {Key Laboratory for Magnetism and Magnetic Materials of the Ministry of Education \& Institute of Theoretical Physics, Lanzhou University, China}

\author{Alexander F. Sch\"affer}

\author{Jamal Berakdar}
\email{jamal.berakdar@physik.uni-halle.de}

\affiliation{Institut f\"ur Physik, Martin-Luther-Universit\"at Halle-Wittenberg, 06099 Halle (Saale), Germany}


\begin{abstract}
Spin waves are the low-energy excitations of magnetically ordered materials. They are key elements in the stability analysis of the ordered phase and have a wealth of technological applications. Recently, we showed that spin waves of a magnetic nanowire may carry a definite amount of orbital angular momentum components along the propagation direction. This helical, in addition to the chiral, character of the spin waves is related to the spatial modulations of the spin wave phase across the wire. It, however, remains a challenge to generate and control such modes with conventional magnetic fields. Here, we make the first proposal for a \textit{magnetic} spiral phase plate by appropriately synthesizing two magnetic materials that have different speeds of spin waves. It is demonstrated with full-numerical micromagnetic simulations that despite the complicated structure of demagnetization fields, a homogeneous spin wave passing through the spiral phase plate attains the required twist and propagates further with the desired orbital angular momentum. While excitations from the ordered phase may have a twist, the magnetization itself can be twisted due to internal fields and forms what is known as a magnetic vortex. We point out the differences between both types of magnetic phenomena and discuss their possible interaction.
\end{abstract}

\maketitle
\section{Introduction}
{ 
Over the past few decades, it has been established that waves with an appropriate spatial structure of the phase can carry a definite amount of orbital angular momentum (OAM). This has been shown for classical  electromagnetic waves as well as for the quantum dynamics of  electrons,  or neutron  (cf. Refs. \cite{photon-vortrex1,photon-vortrex2,e-vortrex1,e-vortrex2,n-vortrex,pyhrep17} and references therein). Also acoustic vortices { have} been recently discussed \cite{PRB99}.

Our  focus here is on magnetic materials and their excitations that propagate within the material itself. To appreciate the difference to electromagnetic waves carrying OAM, for instance, let us recall that the magnetization $\textbf{M}$ (the order parameter in this case), enters the Maxwell-equation indirectly through the constitutive equation $\textbf{B}/\mu_0=\textbf{H}+\textbf{M}$ which is a material-dependent equation ($\textbf{B}$ is the magnetic flux density, and $\textbf{H}$ is the magnetic field strength). The question of interest  for magnetism  is which ground state configuration  is attained by $\textbf{M}$, and what is the structure of  the  excitation spectrum and the corresponding  excited state for a given { sample}. Obviously, it is imperative to set up the free-energy density of the field-free magnetic system, as well as the equation of motion for  $\textbf{M}$. 
Once established, we may in principle couple the magnetic dynamics of $\textbf{M}$ to the Maxwell dynamics  of $\textbf{B}$ (via the above constitutive equation) which allows investigating the photonic and the optomagnetic  properties of the sample, as discussed in the first attempt in Ref.\cite{ramaz}. \\
 The aim here is to solely study $\textbf{M}$ and its excitations in micro-structured magnets. To this end, we start from a free-energy density that accounts for the exchange, the dipolar interactions, the magnetic anisotropies (subsumed in the demagnetization fields) and determine the magnetic configuration as the minimum of this free energy.  The dynamics is governed by the Landau-Lifshitz-Gilbert (LLG) equations \cite{Landau:1981qm} with 
the low-energy eigenmode  excitations being transverse spin waves and with the quanta of these excitations being magnons.
These  collective excitations live only in the ferromagnet and have no signature in the (vacuum) far zone.
Experiments have established that  magnons can be thermally or  non-thermally excited, confined, spectrally shaped, and guided via material  design  and/or microstructuring of the sample \cite{Chumak:2015fa,Hillebrand2,Sadovnikov:2015hm,Sadovnikov:2017hb,Sadovnikov:2018gk,Kruglyak,Vogt,Chumak}.  
Magnonic (spin wave) currents imply  a flow of a  \textit{spin} angular momentum and  are routinely generated at low (thermal) energy cost. Electronic excitations can be regarded as practically frozen at this low energy scale and hence the spin wave propagation is not hindered by Ohmic losses, albeit magnetic losses may still be present and are  described in our scheme collectively by the Gilbert damping. 
A further attractive feature of magnonic-based information processing is the possibility of realizing well-controllable   magnonic logic circuits
\cite{Chumak:2015fa,Hillebrand2,Sadovnikov:2015hm,Sadovnikov:2017hb,Sadovnikov:2018gk,Kruglyak,Vogt,Chumak}
that can be integrated in (spin-)electronic devices. 
Generally,  ferromagnetic spin waves are always right-handed circular polarized waves.  Hence, the use of the polarization state of such waves is somehow restricted.  In view of the markedly different structure of the material and sample-specific   equations of motion for magnetization, as well as those of the Maxwell or the Schr\"odinger equations, it is not clear whether spin waves with OAM may exist as eigenmodes in { ferromagnets} and have  similar properties  as optical or electronic waves with OAM. 
 Recently, we showed under which conditions twisted (helical) magnon beams, meaning OAM-carrying spin waves, do appear as magnetic eigenmode excitations in a waveguide. The amount of OAM along the propagation direction carried  by a  propagating  spin wave is unbound. Such spin waves interact with an external electric field  via the Aharanov-Casher-effect \cite{Aharonov:1984db}, and exhibit an  OAM-controllable magnonic Hall effect \cite{Nakata:2017cd,Onose:2010co}. 
 Despite the differences between magnon and photonic twisted beams which  are discussed in \cite{Jia:2019},  there are several  formal analogies  that can be drawn which should be of a key role for understanding the interaction between twisted magnon and twisted photons, an issue not discussed in this paper. 
 The purpose here is to discuss how to generate and steer twisted magnonic beams. In \cite{Jia:2019}  we applied  at one end of the magnonic waveguide an external  magnetic field that carries OAM.  
 Such fields might be difficult to realize in a general setting.  Thus, we inspect here the possibility of  realizing  a magnonic spiral phase plate (SPP), similar to the one in optics, but now the SPP  is molded out of magnetic materials that have different but (for our purpose) appropriate speed of spin waves. This idea is motivated by the similarity of the simplified magnonic wave equation in a wire  that we used in \cite{Jia:2019}  and the optical waveguide equation. The analytical approach of  \cite{Jia:2019} does not apply however for magnonic SPP, as demagnetization fields and the local magnetic order  are expected to have a strongly  anisotropic spatial distribution, as indeed confirmed below. Hence, we have to employ the  micromagnetic equations  in full generality and resort to  numerical simulations of the spin dynamics in SPP  that we constructed out of  a conducting { permalloy (Py)} and an insulting { yttrium iron garnet (YIG)}.  The results show indeed how a spin wave with a spatially homogeneous phase structure attains OAM when traversing the SPP, and it propagates further as a twisted beam.
 Our numerical method allows at the same time to calculate in addition to the excitations, the thermodynamically stable magnetic states. Interestingly, if the magnetic waveguide is compressed in length to just a disc, magnetic vortices appear as states of the stable magnetizations. Hence, it is of interest to explore 
  how the  geometric confinement-induced magnetic vortices relate to the twisted spin waves when the length of the waveguide becomes extended. This issue is an important first step towards a possible manipulation of magnetic vortices with twisted magnons and is being addressed below. }

{
\section{Magnonic spiral phase plate for generating twisted magnon beams}
	\begin{figure}[bth]
		\includegraphics[width=1\linewidth]{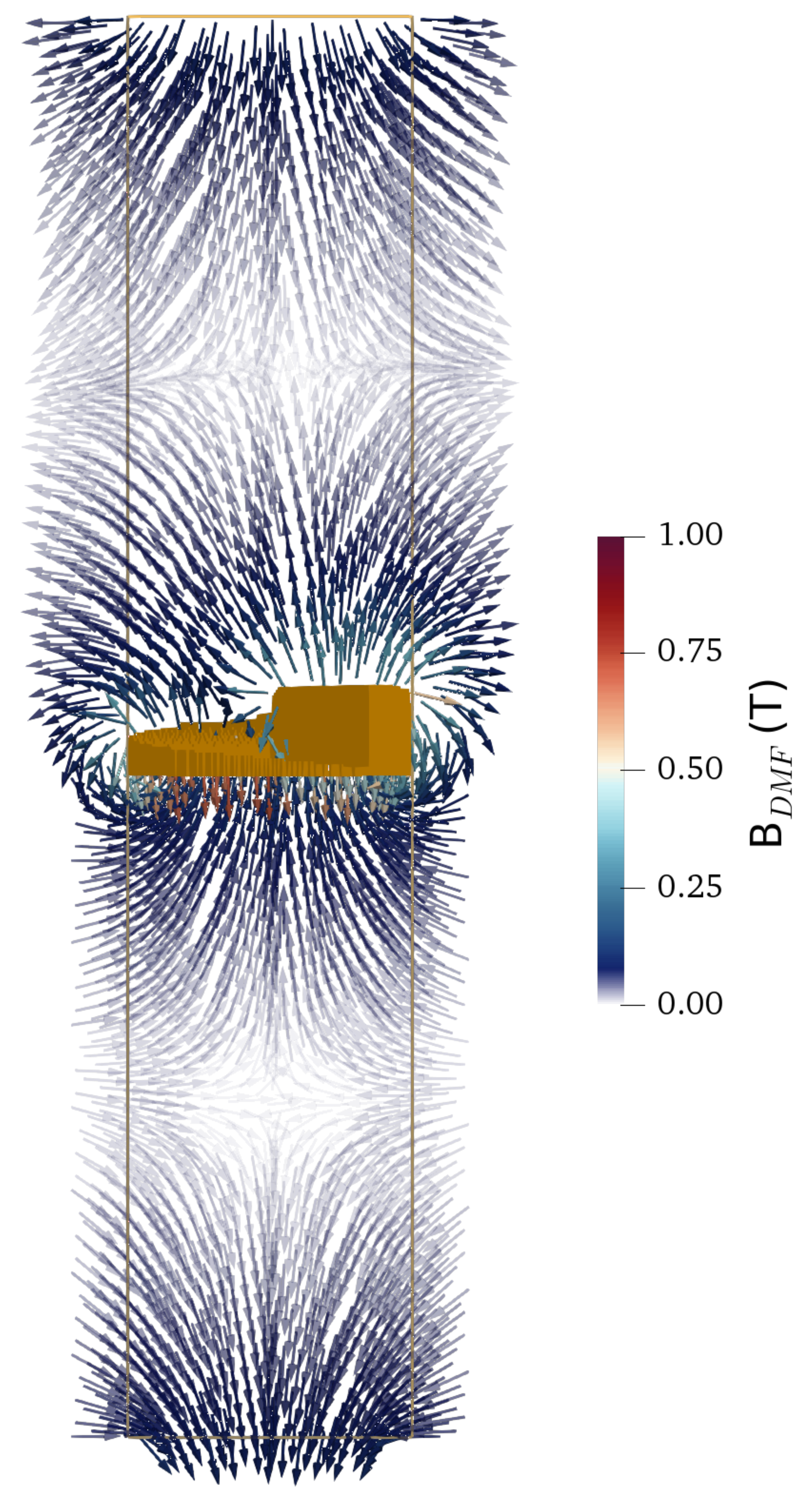}
	\caption{(color online) 
	Demagnetizing field of a cylindrical YIG nanowire (lines mark the wire geometric boundaries) equipped with a SPP (orange structure in the center). The color code corresponds to the magnitude of the demagnetizing fields $\vec{B}_{DMF}$. SPP height: 120~nm. 
	Cylinder radius: $200\,$nm, length: $2000\,$nm. }
	
	\label{Fig::1}
	\end{figure}

	One well-documented  method for generating   optical vortex fields is to use a spiral phase plate \cite{photon-vortrex1}.
	Thus, the question arises of how to construct a magnonic spiral phase plate. 
	 Previously \cite{Jia:2019}  we employed the magnetic component of a spatially structured electromagnetic wave to excite twisted magnon modes in a ferromagnetic waveguide. In contrast,  upon traversing a  magnonic spiral phase plate, a conventional, spatially homogeneous spin wave is transformed into a twisted magnon beam. 
	 
		In order to tackle this challenge, it is expedient to start from the fundamentals of magnetization dynamics.   
		A classical magnetic dipole moment $\vec{m}$ exposed to an external magnetic field $\vec{H}$ precesses around the orientation of the field, a dynamic governed by the equation of motion (EOM)
		\begin{align}
			\dot{\vec{m}}= -\gamma \mu_0(\vec{m}\times\vec{H})~,
		\end{align}
		where $\gamma_0=1.76\times 10^{11}($T$^{-1}$s$^{-1})$ is the gyromagnetic ratio and $\mu_0=4\pi 10^{-7}$VsA$^{-1}$m$^{-1}$ is the vacuum permeability. 
		In practice, magnetic losses result in the  magnetic moments being eventually aligned along the applied magnetic field. This effect can be introduced in the EOM via a phenomenological (Gilbert) damping parameter  $\alpha$  leading  to the  Landau-Lifshitz-Gilbert (LLG) equation \cite{Landau:1981qm} 
		\begin{align}
		\dot{\vec{m}}=-\frac{\gamma \mu_0}{1+\alpha^2}\left[\vec{m}\times\vec{H}+\alpha\vec{m}\times\left(\vec{m}\times\vec{H}\right) \right]~.
		\label{Eq::LLG}
		\end{align}
		Due to the newly established damping term, the magnetization dynamics is defined by a non-linear partial differential equation. 
		By proceeding with a system of interacting magnetic moments which also includes dipolar interactions, the equation is not only non-linear but also non-local as every magnetic moment interacts with all others over their dipolar fields. These stray fields are commonly called demagnetizing fields as their contribution to the free energy functional favors vortex like configurations that diminish the net magnetization of the system. As an example, the demagnetizing field of a YIG cylindrical waveguide including a Py SPP is shown in figure~\ref{Fig::1}.  
		
		In consequence the EOM of a magnetic material differs strongly from the  Maxwell dynamics for electromagnetic waves. 
		Nevertheless, the complex magnetization dynamics still can host spin waves with helical wave fronts as we showed recently \cite{Jia:2019}. These twisted spin waves show analogies to the optical case. 
		The question here is, whether it is also possible to transfer the principle of a spiral phase plate to the magnonic world, in spite of the non-local fields and non-linear dynamics.

	 To assess the feasibility of this basic idea, we set up and conduct micromagnetic simulations using the GPU-accelerated, open-source software package \texttt{mumax3} \cite{VaLe2014}. We solved the LLG equation in a discretized waveguide 
	for every simulation cell $\vec{m}_i$ of the discrete magnetization vector field.
	The time- and space-dependent effective magnetic field,
	\begin{align}
	\vec{B}_i^{\mathrm{EFF}}(t)=\vec{B}_i^\mathrm{EXT}+\vec{B}_i^\mathrm{EXCH}+\vec{B}_i^\mathrm{A}+\vec{B}_i^\mathrm{DMF}~, \label{eq:heff}
	\end{align}
	includes  the conventional external field $\vec{B}_i^\mathrm{EXT}$ that excites spin waves in one end of the waveguide. The  exchange interaction field  is $\vec{B}_i^\mathrm{EXCH}=2A\ind{ex}/M\ind{s}\Delta\vec{m}_i$, with $A\ind{ex}$ being the exchange stiffness and $M\ind{s}$ the saturation magnetization. The uniaxial magnetocrystalline anisotropy field is $\vec{B}_i^\mathrm{A}=2K\ind{u}/M\ind{s} m_z\vec{e}_z$, with $K\ind{u}$ being the anisotropy constant.
	The demagnetizing field reads $\vec{B}_i^\mathrm{DMF}=M\ind{sat}\hat{\vec{K}}_{ij}*\vec{m}_j$, details on the calculation of the demagnetizing kernel $\hat{\vec{K}}$ can be found in \cite{VaLe2014}.
	
	The external magnetic  field, which  is   circularly polarized and is in the radio-frequency range,
	acts on the bottom end of the ferromagnetic wire.
	The peak  amplitude is set to $B\ind{max}=10\,\mathrm{mT}$, the frequency is $\omega=2\pi\times 5\,\mathrm{GHz}$ and the polarization is  circular right-handed.
	
	Upon switching on the external magnetic field magnonic plane waves  are launched.  The task is to convert this wave into a magnon OAM.  To construct a magnonic SPP, we use  two different materials. As in  \cite{Jia:2019,Papp} we use a  wire of  YIG in which the OAM magnonic wave will propagate. In the middle of the wire, the Py SPP nanostructure is inserted. Both materials are widely used in magnonics and have low damping coefficient. Still, the material parameters differ, which should lead to a thickness dependent phase accumulation when passing through a spiral spatial structure.
	
	As a first approach, we disregard the computationally expensive demagnetizing field and check whether a SPP leads to the desired effect of thickness-dependent phase accumulation.
	The resulting wave fronts can be observed  in figure~\ref{Fig::2}. 
	For the presented   results of the simulations, a system of $100\times 100\times 300\,$unit cells (u.~c.) each of $(4\,\mathrm{nm})^3$ size was initialized as a cylindrical base geometry. The Py phase plate (orange structure in figure~\ref{Fig::2}) has a height of 30 unit cells. The YIG region itself is transparent in the graphical representation to be able to show the equal phase fronts of  $m\ind{x}$ instead (red: incident plane wave; blue: transformed helical wave after passing through the SPP).
	
	Depending on the rotational sense of the phase plate, the resulting handedness of the helical wave fronts can be tuned, as shown in figures~\ref{Fig::2}a (counterclockwise) and b (clockwise). Still, for both cases, the polarization of each magnetic moment is right-handed, which is the only possible rotational sense in ferromagnetic materials. 
	It is remarkable that even though the nanostructured SPP is limited by the simulation grid's resolution and hence far from a perfectly smooth phase plate, the resulting twisted magnon wave shows the characteristics of the theoretically proposed OAM carrying magnon beam.
	
	In addition to the described basic geometry,  the system outer boundaries are modeled to absorb most of the incoming magnons by introducing an exponentially increasing damping parameter in a tube shell with a $20\,$u.~c. thickness. This helps to avoid reflection effects.
	The used material parameters for the YIG wire are: $M\ind{s}=1.4 \times 10^5 $ A m$^{-1}$, $ A\ind{ex}=3 \times  10^{-12}$  J m$^{-1}$, $K_u=5\times 10^3$  J m$^{-3}$, $\alpha= 0.01$\cite{xiguang}. In the case of Py we assume: $M\ind{s}=8.3 \times 10^5 $ A m$^{-1}$, $ A\ind{ex}=10 \times  10^{-12}$  J m$^{-1}$, $K_u=-1\times 10^3$  J m$^{-3}$, $\alpha= 0.02$\cite{Coey}.
	
	When dealing with ferromagnetic systems on the micrometer scale, usually stray fields play an important role. This also holds for the cylindrical wires, where not only do the outer surfaces of the structure lead to strong demagnetizing fields but also the boundary between SPP and YIG nanowire will have an impact, mainly due to the large difference in the saturation magnetization of both materials. 
	For a correct interpretation of the simulation results, one has to keep in mind that because of the demagnetizing field, the magnetization is not homogeneously aligned anymore but shows a canting close to the surface. Therefore, the excitation of the magnetization in figure~\ref{Fig::3} is calculated relative to the relaxed magnetization before applying an external field. 
	Figure~\ref{Fig::3} shows the $x$ component of the relative excitation after a $\omega=2\pi\times 10~$GHz plane wave has passed the SPP (orange). In addition to the time dependent magnetic field, a 1T static magnetic field   was applied along  the $z$ direction to reduce the magnon wavelength and therefore check the propagation after more periods in the same geometry as before.  A few observations are striking. 
	First, we can identify the region after the SPP with a helical magnon beam, as some singularity in the excitation is present, separating different phases of the wave in its cross-section.
	Second, it is obvious that the region directly adjacent to the SPP shows some rather chaotic patterns for the magnetization excitation. This partially arises due to the limited resolution of the SPP but also is a footprint of the demagnetizing fields hindering the passage of  spin waves (see field distribution in figure~\ref{Fig::1}). Third, there is a transition region where the center of the twisted beam is not coinciding with the center of the waveguide for similar reasons. However, these two symmetry axes align with each other after some distance. 
	
	Summarizing this section, we showed that in spite of the fundamental differences between the nature of magnetization dynamics -- \emph{non-linear}, \emph{non-local} -- and electromagnetic waves in free space, the concept of the SPP for the transformation of plane waves to helical waves holds for the field of magnonics. 
	The long-range dipolar interaction does not necessarily endanger the spatial structure of magnon beams carrying orbital angular momentum but will have an even stronger influence as in the presented example for less symmetric geometries like cuboid-shaped waveguides, which has to be addressed in future studies.	
		
	\begin{figure}[bth]
		\centering
		\includegraphics[width=0.75\linewidth]{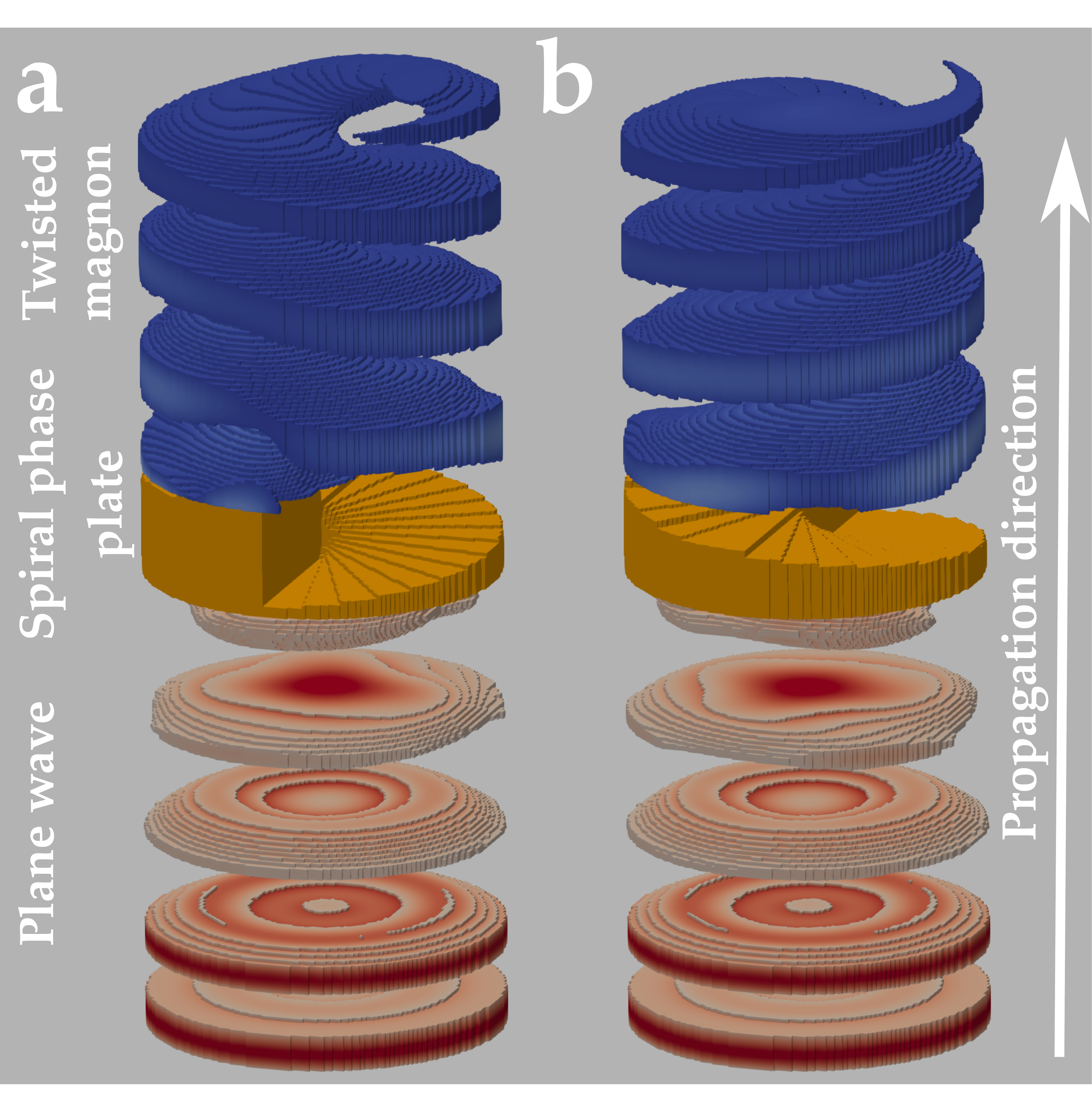}  
		\caption{(color online) 
			The transformation of an incident magnonic plane wave into twisted magnons by a magnonic spiral phase plate disregarding demagnetizing fields. The blue and red regions show wave fronts of equal phases of the $m\ind{x}$ component, clearly demonstrating  the change from planar to helical wave fronts. The orange regions show the geometries of the spiral phase plate generating either counterclockwise \textbf{a} or clockwise \textbf{b} rotations of the twisted magnons. } 
		\label{Fig::2}
	\end{figure}

	\begin{figure}[bth]
		\centering
		\includegraphics[width=1\linewidth]{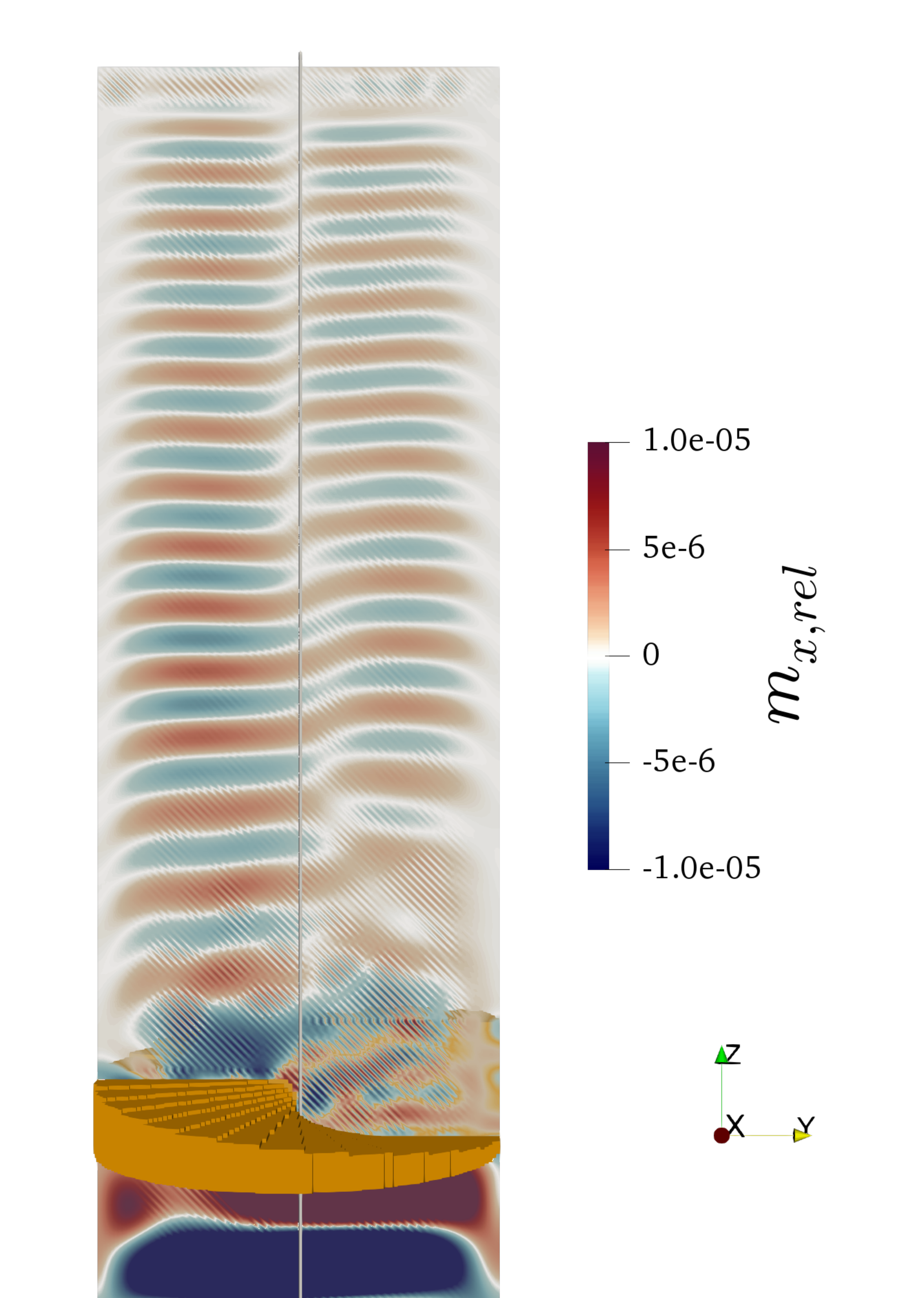}  
		\caption{(color online) 
			The transformation of an incident magnonic plane wave into a twisted magnon by a magnonic spiral phase plate considering demagnetizing fields. 
			The color code corresponds to the excitation of the magnetization in the $yz$ plane relative to the initial relaxed state (see figure~\ref{Fig::1}).
			The blue and red regions show wave fronts of equal phases of the $m\ind{x}$ component, demonstrating  the change from planar to helical wave fronts. The orange regions show the geometry of the spiral phase plate. 
			} 
		\label{Fig::3}
	\end{figure}

}	

\section{Magnetic vortex versus twisted magnons}

\begin{figure}[t]
	\centering
	\includegraphics[width=\linewidth]{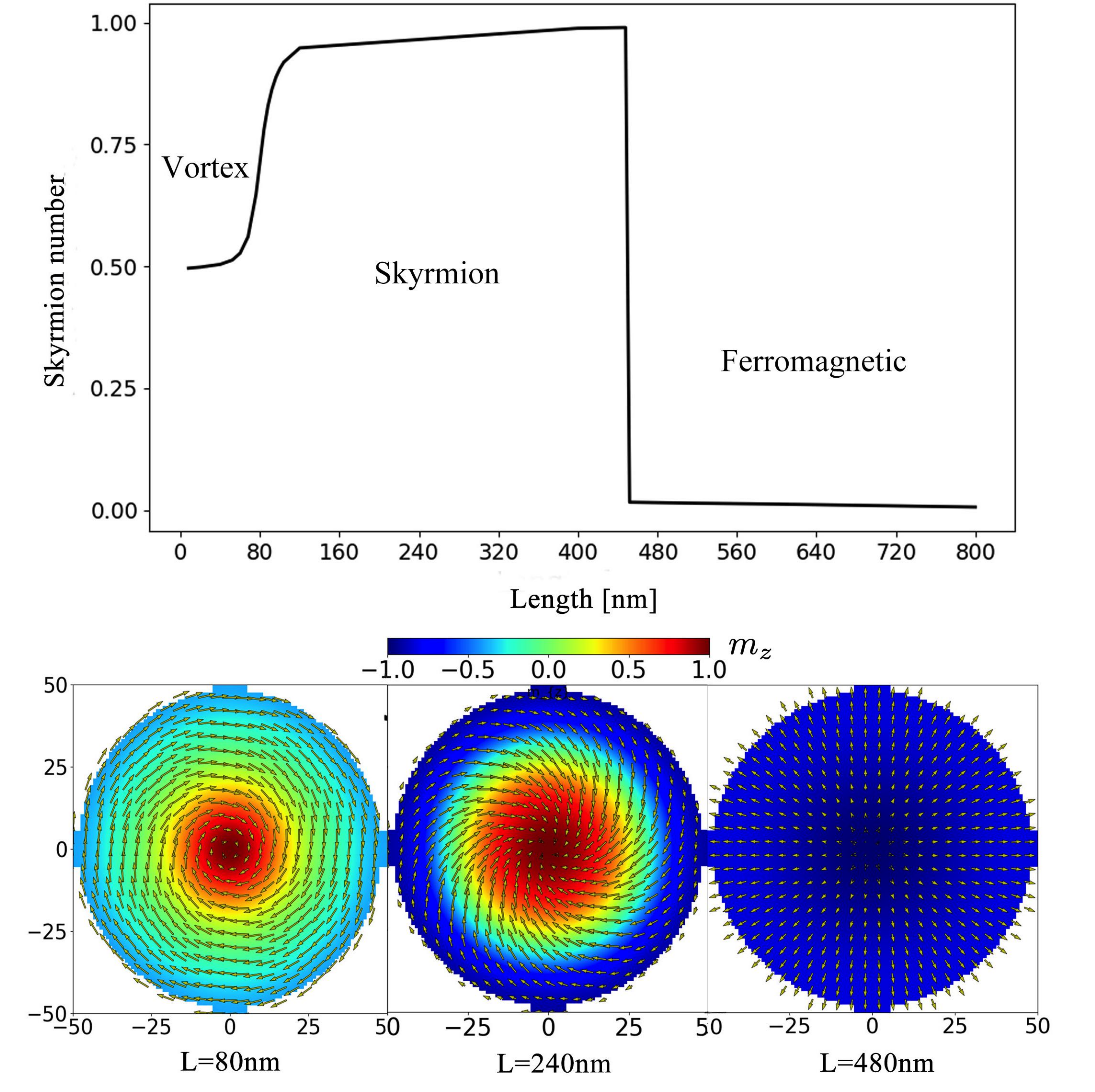}  
	\caption{(color online) 
		(Upper inset) The evolution of the magnetic configuration (as characterized by the skyrmion number) in YIG waveguides  with varying length $L$ and a fixed radius of $R=200$ nm.  (Lower inset) Slices showing the typical configurations of the stable magnetization  state with increasing length $L$: a magnetic vortex, a magnetic skyrmion, and a ferromagnetic state. } 
	\label{Fig::4}
\end{figure}

\begin{figure}[t]
	\centering
	\includegraphics[width=\linewidth]{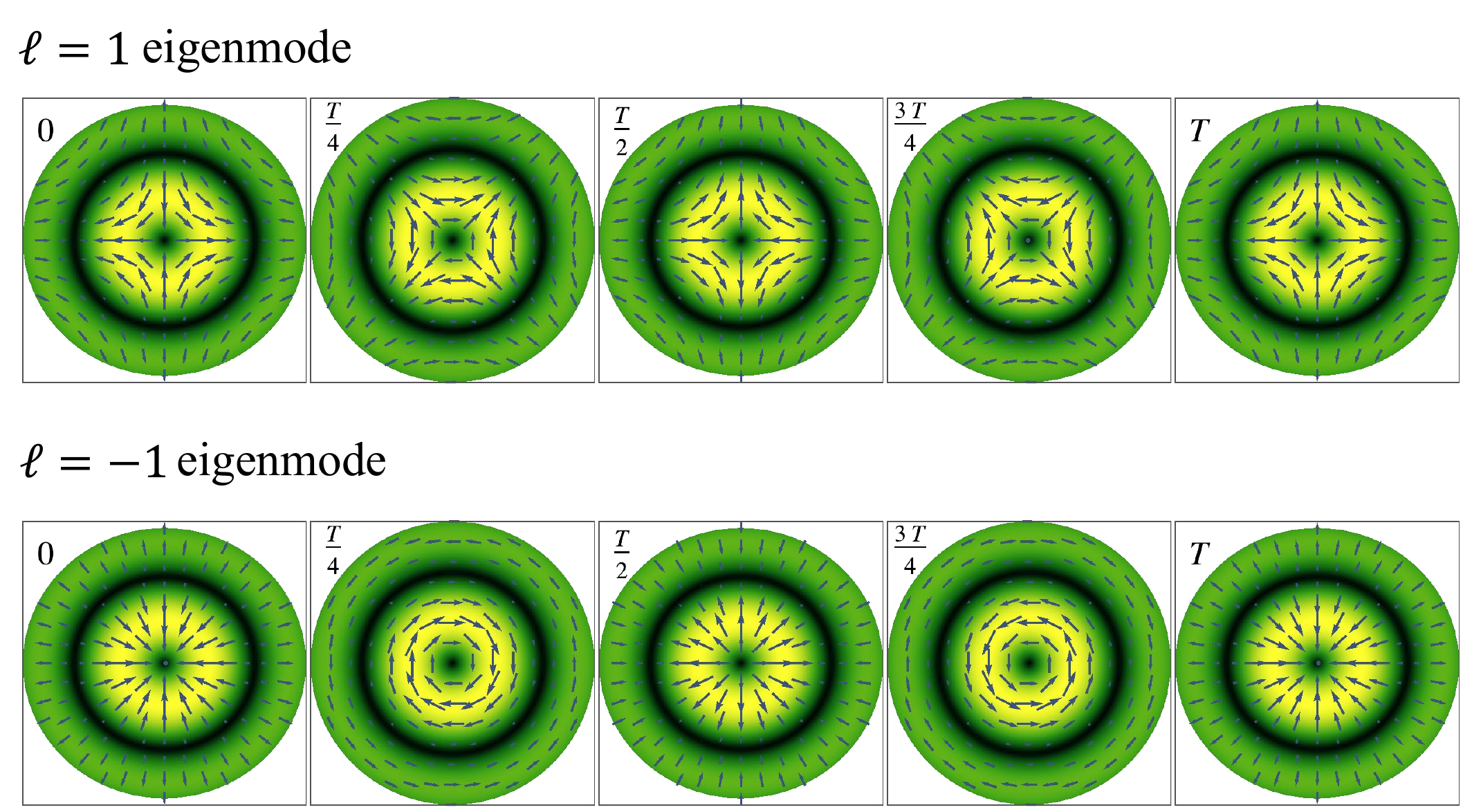}  
	\caption{(color online) 
		The dynamic spin configurations. Snapshots showing the topological difference between the magnetization vortex \cite{Wirtz} and twisted magnons: The chirality of vortices is very robust by virtue of the  rotational symmetry and strong exchange interaction. In contrast, the chirality of twisted beams varies  with the radius and/or the time. Here, $T$ is the period of magnetization precession. 
		With the  boundary conditions of vanishing magnetic fluctuations at the boundary and outside the magnetic sample, one can have a  square wave oscillation of the chirality with  time $t$ and/or  distance $z$.} 
	\label{Fig::5}
\end{figure}

Magnonic twisted beams are  in essence small transverse deviations from the collinearly ordered phase.
Hence, they are qualitatively different from other well-known twisted magnetic states, for instance,  magnetic vortices  which are intrinsically non-collinear in their stable configuration \cite{hertel1,hertel2,hertel3}.
It has been shown  theoretically and experimentally that for thin ferromagnetic disks and other-shaped elements of micrometer size and below, the magnetic vortex is energetically favored. As one kind of topological magnetic excitations, magnetic vortices are  driven by a subtle competition between geometrical balance of exchange interaction and the demagnetizing field (i.e. the shape anisotropy): the magnetization tends to align parallel to the surface in order to minimize the surface charge, however, the singularity at the center of a vortex is replaced by an out-of-plane magnetized core in order to reduce the exchange energy.  Magnetic vortices are excited from the uniform ferromagnetic state with much higher energy than the  magnonic twisted eigenmode.  Topologically, a magnetic vortex is characterized by two essential features: the out-of-plane,  nanometer-size vortex-core magnetization. This determines  the vortex \textit{polarity} $p$ which can be up or down.  The second feature is related to the in-plane spin configuration of the vortex (say with extension $r \gg 1$ in scaled units) described by $\vec{m} (r \gg 1, \phi) = \cos \Phi \hat{\vec{e}}_{x} + \sin \Phi \hat{\vec{e}}_{y}$ with $\Phi = q \phi + c \pi/2$.  The  topological \textit{vorticity}  is set by  $q$ indicating  the winding number, where  $c$ is the vortex \textit{chirality}. As shown in figure~\ref{Fig::4}, the stable magnetization of YIG waveguides is simulated with  different lengths $L$. The topology of the magnetic configuration is characterized by the skyrmion number $s =\sum_{\vec{r}} \chi_{\vec{r}}/L$, where $\chi_{\vec{r}}=\vec{m}_{\vec{r}}\cdot(\vec{m}_{\vec{r}+\vec{x}}\times\vec{m}_{\vec{r}+\vec{y}})+\vec{m}_{\vec{r}}\cdot(\vec{m}_{\vec{r}-\vec{x}}\times\vec{m}_{\vec{r}-\vec{y}})$.  For example, a vortex with a winding number $q$ and core polarization $p$ has a half-integer skyrmion number $s=qp/2$, while a skyrmion has integer $s= qp$ and ferromagnetic (FM) state has $s=0$. Clearly, the demagnetizing field plays a crucial role in the formation  of the magnetic configurations. The magnetic vortex states are stable only if the waveguide is compressed into a magnetic disc.
Note that due to rotational symmetry of the magnetic vortex and the strong exchange interaction in ferromagnetic materials, a large energy barrier prevents  changing easily  the chirality of the vortices. For thicker magnetic disks, skyrmions are stabilized by the demagnetizing fields, whereas in the limit of long cylindrical wire the exchange interaction is dominant and leads to a stable  ferromagnetic configuration. 

Turning to  the magnonic twisted eigenmode of the FM waveguide, the topological vorticity is given by the twisted beam OAM, $-\ell$. The skyrmion number is, however, nearly zero.  The chirality is not fixed as well. It depends on the radius $r$, the time $t$, and/or the distance $z$, as demonstrated in  figure~\ref{Fig::5}. Furthermore, it is well-known that a magnetic vortex-antivortex pair is unstable: the pair with parallel polarization belongs to the same topological sector as a uniform FM ground state,  meaning that we can deform the pair continuously into the ground state once the vortex-antivortex pair with parallel polarization was created; whereas, the vortex-antivortex pair with antiparallel polarization has a nontrivial topological number. Its annihilation is accompanied  with the ejection of a magnetic monopole and spin waves. The magnonic twisted eigenmodes with $\pm \ell$ are degenerate states, they can be excited simultaneously and result in a right-handed superposition $|R \rangle = \xi_{+} |+\ell \rangle + \xi_{-} | -\ell \rangle$, which defines a new degree of freedom that  we can refer to as the magnonic vorticity.

As for the interplay between the magnonic twisted beams and a magnetic vortex, it has been shown that a smooth background magnetic texture affects the dynamics of a spin wave by changing the basic equation that rules the spin wave behavior \cite{Santos:2015ej}. An O(2) vector potential and a scalar potential enter in the Schr\"{o}dinger-like equation for the magnons.  The SPP allows us to investigate the scattering of magnonic twisted beams by topologically nontrivial magnetic textures, including magnetic vortices, magnetic skyrmions, and magnetic Bloch points.

\section{Summary}
We discussed {several} aspects of  magnonic beams carrying orbital angular momentum. We highlighted the characteristics of these topological excitations  contrasting them with the well-known magnetic vortices.
For generating OAM carrying magnonic beams, we proposed and realized numerically a magnonic spiral phase plate on the basis of the  YIG/Py heterostructure.   Full-fledged numerical simulations using the heterostructure endorse the experimental feasibility of this type of helical magnonic beams. 
%

\section{Acknowledgements}
This work is supported by the National Natural Science Foundation of China (No. 11474138, and No. 11834005), the German Research Foundation (No. SFB 762, and SFB TRR 227), and the Program for Changjiang Scholars and Innovative Research Team in University (No. IRT-16R35).

\end{document}